\documentclass[conference]{IEEEtran}

\ifCLASSINFOpdf

\else

\fi
\usepackage{tikz}

\newcommand\copyrighttext{%
  \footnotesize 2024 IEEE. Personal use of this material is permitted. Permission from IEEE must be obtained for all other uses, in any current or future media, including reprinting/republishing this material for advertising or promotional purposes, creating new collective works, for resale or redistribution to servers or lists, or reuse of any copyrighted component of this work in other works.}
\newcommand\copyrightnotice{%
\begin{tikzpicture}[remember picture,overlay]
\node[anchor=south,yshift=10pt] at (current page.south) {\fbox{\parbox{\dimexpr\textwidth-\fboxsep-\fboxrule\relax}{\copyrighttext}}};
\end{tikzpicture}%
}

\usepackage{amsmath,amssymb}
\usepackage{graphicx}
\usepackage{xspace}
\usepackage{color}
\usepackage{epstopdf}
\usepackage{algorithm}
\usepackage{algorithmic}
\usepackage{setspace}
\usepackage[font=small,labelfont=bf]{caption}
\usepackage{subcaption}
\usepackage{pgfplots, tikzscale}
\usetikzlibrary{plotmarks}
\usetikzlibrary{plotmarks,patterns}
\pgfplotsset{compat=newest}
\usepackage{multirow}
\definecolor{darkpastelgreen}{rgb}{0.01, 0.75, 0.24}
\definecolor{azure}{rgb}{0.0, 0.5, 1.0}
\usetikzlibrary{decorations.pathreplacing,calligraphy}
\usetikzlibrary{shapes.misc}
\usepackage[hyphens]{url}

\newcommand{\systemname}{DARE\xspace}
\graphicspath{ {figures/} }
\def\thesubsubsectiondis{\unskip\arabic{subsubsection})}
\begin{document}
\bstctlcite{IEEEexample:BSTcontrol}



\title{\fontsize{18.9}{18.9}\selectfont Ultra-Low-Complexity, Non-Linear Processing for MU-MIMO Systems}
\author{\IEEEauthorblockN{Chathura Jayawardena and Konstantinos Nikitopoulos}
\IEEEauthorblockA{5G \& 6G Innovation Centre, Institute for Communication Systems (ICS), University of Surrey, Guildford, UK}}
\maketitle
\IEEEpeerreviewmaketitle
\copyrightnotice
\begin{abstract}

Non-linear detection schemes can substantially improve the achievable throughput and connectivity capabilities of uplink MU-MIMO systems that employ linear detection.
However, the complexity requirements of existing non-linear soft detectors that provide substantial gains compared to linear ones are at least an order of magnitude more complex, making their adoption challenging. In particular, joint soft information computation involves solving multiple vector minimization problems, each with a complexity that scales exponentially with the number of users. This work introduces a novel ultra-low-complexity, non-linear detection scheme that performs
joint Detection and Approximate Reliability Estimation (DARE).
For the first time, DARE can substantially improve the achievable throughput (e.g., $40\%$) with less than $2\times$ the complexity of linear MMSE, making non-linear processing extremely practical. To enable this, DARE includes a novel procedure to approximate the reliability of the received bits based on the region of the received observable that can efficiently approach the accurately calculated soft detection performance.  In addition, we show that DARE can achieve a better throughput than linear detection when using just half the base station antennas, resulting in substantial power savings (e.g., $500$ W). Consequently,
DARE is a very strong candidate for future power-efficient
MU-MIMO developments, even in the case of software-based
implementations, as in the case of emerging Open-RAN systems. Furthermore, \systemname can achieve the throughput of the state-of-the-art non-linear detectors with complexity requirements that are orders of magnitude lower.
\end{abstract}

\section{Introduction}

    
Due to their favorable complexity requirements, existing MIMO developments employ linear (e.g., MMSE) based detectors in the uplink. Still, such detectors leave unexploited throughput and connectivity benefits, making them inefficient in current and next-generation communication systems. 
In particular, the achievable throughput by linear detection is severely degraded when the MIMO channel matrix is poorly conditioned \cite{nikitopoulos2014geosphere,Antipodal,gmultisphere}. For example, such MIMO channels occur when the number of transmitted streams is close to the number of receiver antennas.
This limitation has led to massive MIMO systems where the base station antennas are much larger than the supported number of streams. However, a massive number of base station antennas (e.g., 128) requires a massive number of RF chains that consume excessive power to support a relatively small number of streams (e.g., 12). 



 In contrast, non-linear detection can overcome the limitations imposed by linear detection and provide substantial throughput and connectivity gains compared to linear detection \cite{nikitopoulos2018massively,gmultisphere}. Furthermore, in Section \ref{s:Eval}, we elaborate that non-linear detection can deliver better throughput than linear detection while significantly reducing the number of base station antennas and, therefore, RF chains. As a result, the power consumption of a base station can be reduced substantially by employing non-linear processing. 
 To achieve these gains, non-linear processing schemes that can accurately compute soft information are necessary, leveraging channel decoding schemes employed in current standard-based systems.
 However, the complexity requirements of the joint soft information computation (e.g., in the form of Max-Log optimal Log-Likelihood Ratios (LLRs) \cite{STS}) are still substantially higher than linear processing for a large number of concurrently transmitted streams \cite{newsoftfsd,SFSD,STS}. 
 This is because joint soft information computation typically involves solving multiple minimization problems, each with a worst-case complexity that scales exponentially with the number of users. 
 Many approximate non-linear hard detection schemes exist. However, the processing complexity requirements per received vector sample of schemes that are based on message passing algorithm \cite{CHEST}, local neighborhood search \cite{ULAS}, and convex optimization \cite{lowdet} 
  is of the order $O(K^2M)$, where $K$ is the number of users and $M$ is the number of base station antennas. Recently proposed deep learning-based GEPNeT detection scheme can approach optimal hard detection performance \cite{GEPNET}.  However, the complexity order of GEPNeT exceeds $O(K^2M)$ without even considering the training phase \cite{GEPNET,carlos_deep_2023}. These requirements
  are substantially higher than $O(MK)$ of MMSE, even for performing hard detection. Furthermore, the achievable performance of schemes such as \cite{ULAS} significantly degrades when transmitting dense symbol constellations (e.g., 64 QAM). In addition, the performance of message passing \cite{CHEST,VAMP} is highly dependent on the statistics of large systems. 
 
 Tree search-based sphere decoders (SD) are promising to achieve the optimal hard Maximum Likelihood (ML) \cite{nikitopoulos2014geosphere} performance and Max-Log optimal soft detection performance \cite{STS} in the MIMO uplink. All SD schemes consist of a channel matrix-dependent preprocessing stage and a per-received vector post-processing stage. The channel matrix-dependent preprocessing stage involves a triangular (e.g., QR) decomposition and is only required to be performed once the channel changes significantly, similar to the inversion of linear detection, and with similar complexity requirements. However,  the complexity requirements of the per-received vector post-processing stage in SD schemes are many orders of magnitude higher than that of linear processing \cite{FCSD,SFSD,newsoftfsd}.
 
 The recently introduced massively parallelizable non-linear (MPNL) detection scheme has been shown to be efficient \cite{gmultisphere} in approaching optimal performance and capable of outperforming state-of-the-art detectors. MPNL detection scheme can minimize processing latency while achieving  ML performance by dividing the detection process into parallel processes that do not interact. In contrast, we exploit dependencies in this work to maximize performance gains specifically for a smaller complexity.
 
 In the context of low-complexity non-linear detection, the antipodal detection and decoding principle introduced in \cite{Antipodal} can outperform existing non-linear detection schemes with a better tradeoff between performance and complexity. However, current Antipodal approaches rely on statistics of large systems and become approximate for a smaller number of streams. Furthermore, the current Antipodal approach accepts or discards the whole vector. Therefore, it may discard symbol (and therefore bit) estimates of all users as unreliable for a particular realization, while some symbols (and therefore bits) (e.g., of strong users) could still be reliable. 

   With the popularity of the open-RAN paradigm,  power-efficient physical layer solutions that can enhance network performance are timely and necessary. Such solutions are required to meet the stringent latency requirements of the 3GPP physical layer, even in a softwarerized implementation. Therefore, a non-linear detection scheme that can substantially reduce the power consumption of a base station with ultra-low complexity requirements becomes an ideal candidate for modern physical layer developments. A practical non-linear detection scheme must ideally have a fixed latency and complexity and be capable of delivering substantial gains compared to linear detection, with a very small complexity increase. A comparable small complexity increase (e.g., $< 2\times$) can enable the exchange of linear detection with non-linear detection in exiting deployments without significant modifications to the architecture and without compromising the supported bandwidths and the number of user streams.

 
 In this work,  we introduce Detection and Approximate Reliability Estimation  (\systemname):, a novel highly-efficient ultra-low-complexity non-linear detection scheme. \systemname can achieve near-optimal hard ML and soft detection performance \cite{STS} with a time complexity order of $O(MK)$ per received vector sample.  
 To enable this, \systemname exploits a novel detector structure to provide soft bit reliability information based on the region of the received observable (Section \ref{s:method}). Consequently, \systemname can approximate the optimal soft information computation with lower complexity than exiting non-linear detectors that provide hard estimates. \systemname can efficiently quantize the reliability information as a function of complexity, providing a flexible performance/complexity tradeoff.  Furthermore, \systemname can compute reliability information in a hardware-friendly manner while avoiding any sorting operations, which is a bottleneck for existing non-linear detectors \cite{Shabany08,NewKbest}. In contrast to the Antipodal approach, \systemname applies to a smaller number of streams, determines the reliability of bits per user basis (and does not characterize the whole vector), and has a fixed processing latency. As a result, for the first time, \systemname can significantly outperform linear soft detection (e.g., throughput gains of 40\% even in massive MIMO scenarios) with a maximum complexity that is only $2 \times$ than linear detection (Section \ref{s:Eval}). 
Furthermore,  \systemname can provide better throughput than linear MMSE while employing half the base station antennas, resulting in power savings of $500$W \cite{RFpow} for a $64$-antenna base station.

\section{System Model}
A spatially multiplexed uplink Multi-User MIMO system with $K$ single-antenna users transmitting to an $M$-antenna base station is assumed. Then, the complex baseband model is given by 
\begin{equation}
   \textbf{y = Hs+n}  
\end{equation}
 where $ \textbf{y} $ is the $M \times 1 $ received vector, $\mathbf{s}$ is the $K\times1$ transmitted symbol vector with elements belonging to a constellation $\mathcal{O}$, $\textbf{H}$ is the $M \times K$ channel matrix, and $\textbf{n}$ is the $M \times 1$ an additive white Gaussian noise vector with variance $\sigma^2$.

\section{Multi-Layer Joint Processing for MU-MIMO}
In practical systems that employ soft channel decoding approaches like LDPC, soft information is required in the form of Log Likelihood Ratios (LLRs). The LLR for the $jth$ coded bit $b_j$ is defined as in \cite{STS,softAPPmimo} 
\begin{equation}
L(b_{j}) \triangleq  \ln{\bigg(\frac{P[b_{j}=+1|\mathbf{y,H}]}{P[b_{j}=-1|\mathbf{y,H}]}\bigg)}.
\end{equation}
The computation of LLRs, when the Max-Log approximation is employed, involves multiple constrained ML searches \cite{STS, softAPPmimo}. In particular, the LLR for the $j$th coded bit $b_j$ could be expressed as
 \begin{align}
\nonumber L(b_j) \approx &  \min_{\mathbf{s}\in S_{j}^{ -1}}\bigg\{\frac{1}{\sigma^2} \|{\textbf{y}} -{\mathbf{H}}\textbf{s}\|^2\bigg\}- \min_{\mathbf{s}\in S_{j}^{ +1}}\bigg\{ \frac{1}{\sigma^2}\|{\textbf{y}} -{\mathbf{H}}\textbf{s}\|^2\bigg\}\\
 = & \text{sign}(x_j) (D_j^{\overline{ML}}-D^{ML}), 
 \label{eq:LLR}
\end{align}
where  $D_j^{\overline{ML}}=\min_{\mathbf{s}\in S_{j}^{\bar{x}_j}}\bigg\{ \frac{1}{\sigma^2}\|{\textbf{y}}-{\mathbf{H}}\textbf{s}\|^2\bigg\}$, $D^{ML}=\min_{\mathbf{s}\in 
  \mathcal{O}^{M}}\bigg\{ \frac{1}{\sigma^2}\|{\textbf{y}}-{\mathbf{H}}\textbf{s}\|^2\bigg\}$ and $x_j$ is the $j$th entry of the ML solution's bit label and $S_{j}^{ -1},S_{j}^{ +1}$ are the subsets of possible symbol vectors with $j$th bipolar bit set to $-1,+1$ respectively. Here $D^{ML}$ is the metric of the ML solution and $D_j^{\overline{ML}}$ is the minimum metric from subset $S_{j}^{\bar{x_j} }$ for bit $j$. 
We note that the LLR calculation in Eq. (\ref{eq:LLR}) is of impractical complexity to compute optimally for a larger number of layers and modulation orders. The next Section introduces an improved detector that can well approximate this LLR calculation with low complexity requirements.

\section{Detection and Approximate Reliability Estimation}
\label{s:method}
This Section describes the design of \systemname together with its complexity analysis. Section \ref{s:Apprx} discusses the preprocessing (QR decomposition) and 
Section \ref{s:PervecP} introduces the details of per vector processing of \systemname. In particular, \systemname includes a novel procedure to identify unreliable bits based on the region of the received observable. Then, the candidate symbols corresponding to these bits are selected from all layers and updated by a new tree traversal strategy. Finally, a refined bit reliability estimate is obtained based on the selected candidates. This estimate can well approximate the LLR calculation in Eq. (\ref{eq:LLR}) for an increasing number of candidates considered. 
Finally, the complexity of \systemname is discussed. 

\subsection{Preprocessing (QR decomposition)}
\label{s:Apprx}
To transform the reliability estimation problem into a tree search, the channel matrix  $\mathbf{H}$ is decomposed to an orthonormal $\mathbf{Q}$ and upper triangular $\mathbf{R}$ matrices. 
The generalized QR decomposition of a Tikhonov regularized matrix $\bar{\mathbf{H}}$, can be defined as
\begin{equation}
\bar{\mathbf{H}}\triangleq\begin{bmatrix} \mathbf{H} \\ \lambda\mathbf{I}_{K} \end{bmatrix} = \bar{\mathbf{Q}}\mathbf{R}=\begin{bmatrix} \mathbf{Q} \\  \mathbf{Q}_2 \end{bmatrix} \mathbf{R}, 
\label{eq:reg}
 \end{equation}
with the regularization parameter $\lambda=\sigma/\text{E}\{|s_l|\}$ and $\text{E}\{|s_l|\}$ denoting the expected symbol energy. This regularization can reduce the impact of channel estimation error at low SNRs and also mitigate the effect of an ill-determined ${\mathbf{H}}$ matrix \cite{gmultisphere}. For example, the  ${\mathbf{H}}$ matrix can be ill-determined due to the spatial correlation of antennas and/or users. The  ${\mathbf{H}}$ matrix also becomes  ill-conditioned when the number of users approaches the number of base-station antennas. 
Here, $ \mathbf{Q}$ is a $M\times K$ matrix and $\mathbf{R}$ is a $K\times K$ upper triangular matrix.

\subsection{Per received vector processing}
\label{s:PervecP}
\noindent The per vector processing of \systemname consists of a \textit{candidate selection} procedure which identifies the symbols corresponding to unreliable bits based on the region of the received observable. Then, these selected candidates are updated starting from root layer $K$ to leaf layer $1$ by  a \textit{tree traversal and candidate update} strategy. Lastly, bit  reliability estimates are obtained from the selected candidates by a \textit{reliability estimate computation} procedure, which distinguishes the unreliable and reliable bits and approximates the corresponding reliability estimate. The following paragraphs describe these three procedures, while the Pseudocode in Algorithm \ref{pc:sortfreeARE} further describes the steps in detail.

\noindent First, the effective $K \times 1 $ received observable vector is obtained by
\begin{equation}
\mathbf{\tilde{y}}=\mathbf{Q}^H\mathbf{y}
 \end{equation}
Then, the received observable at level $K$ is given by 
\begin{equation}
\hat{y}_{K}=\frac{\tilde{y}_{K}}{R_{K,K}}
 \end{equation} 

 \noindent \textit{Candidate selection:} Next, based on the position of $\hat{y}_{K}$ relative to the constellation points, the candidate symbols considered for the layer $K$ are determined, which results in the reliability estimation of the corresponding bits. 
 \begin{figure}[!h]
	\centering
	\includegraphics[width=0.45\columnwidth]{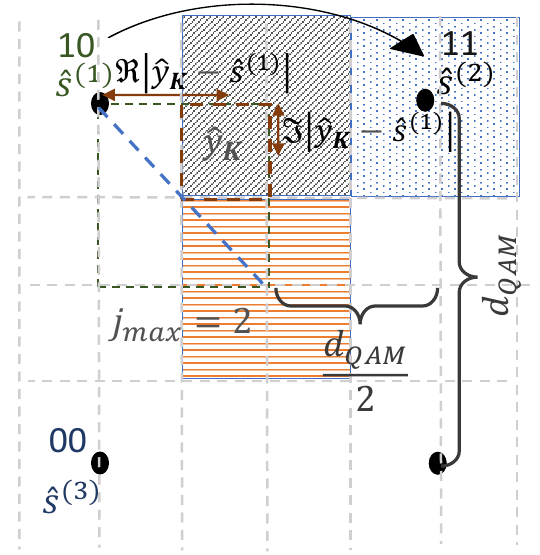}
	
	\caption{Candidate selection example for $l=K,{N}_{C}=4$.}
	
	\label{f:enum}
 \vspace{-10pt}
\end{figure}

 To attain this, candidate symbols closest to $\hat{y}_{K}$  and their ordering needs to be identified  in a computationally efficient manner. 
 To avoid the computation and sorting of $|\mathcal{O}|$ candidate symbol distances, the geometry of the constellation can be exploited to identify the $j_{max}$ symbols closest to $\hat{y}_{K}$.  
 
 Simply, in Fig. \ref{f:enum} based on the position of $\hat{y}_{K}$ the first closest symbol $\hat s^{(1)}$, the second closest symbol $\hat s^{(2)}$ and the third closest symbol can be identified. In particular,  $\hat s^{(1)}$ can be identified by slicing $\hat{y}_{K}$  on the constellation decision boundaries ($\hat{s}^{(1)}=\lfloor\hat{y}_{K}\rceil$). Then, $\hat s^{(2)}$ and $\hat s^{(3)}$ can be determined based on $\Re|\hat{y}_{K}-\hat{s}^{(1)}|\lessgtr\Im|\hat{y}_{M}-\hat{s}^{(1)}|$. This is explained in the Pseudocode of Algorithm \ref{pc:ordering} for a general $\hat{y}_{l,i}$. 
 Pseudocode \ref{pc:ordering} exploits the initial steps in the two-dimensional zigzag method introduced in \cite{nikitopoulos2014geosphere} to determine the ordering of the first four closest symbols. Here $\lfloor \rceil$ is required when $\hat{s}^{(1)}$ is at the edges of the constellation.
 \begin{algorithm}[h] 
	\caption{ Pseudocode for the symbol ordering}
	\label{pc:ordering}
	{\fontsize{8.5}{8.5}\selectfont
	\begin{algorithmic}[1]
		\IF{$\Re|\hat{y}_{l,i}-\hat{s}^{(1)}|> \Im|\hat{y}_{l,i}-\hat{s}^{(1)}|$}
			\STATE{$\hat{s}^{(2)}=\lfloor\text{sign}(\Re(\hat{y}_{l,i}-\hat{s}^{(1)}))d_{QAM}+\hat{s}^{(1)}\rceil$}
			\STATE{$\hat{s}^{(3)}=\lfloor j\text{sign}(\Im(\hat{y}_{l,i}-\hat{s}^{(1)}))d_{QAM}+\hat{s}^{(1)}\rceil$}
			\ELSE
		    \STATE{$\hat{s}^{(2)}=\lfloor j\text{sign}(\Im(\hat{y}_{l,i}-\hat{s}^{(1)}))d_{QAM}+\hat{s}^{(1)}\rceil$}
			\STATE{$\hat{s}^{(3)}=\lfloor\text{sign}(\Re(\hat{y}_{l,i}-\hat{s}^{(1)}))d_{QAM}+\hat{s}^{(1)}\rceil$}
			\ENDIF
	       \STATE{$\hat{s}^{(4)}=\lfloor\text{sign}(\Re(\hat{y}_{l,i}-\hat{s}^{(1)}))d_{QAM}+j\text{sign}(\Im(\hat{y}_{l,i}-\hat{s}^{(1)}))d_{QAM}+\hat{s}^{(1)}\rceil$}
	
 	\end{algorithmic}}
\end{algorithm}
 
 The maximum number of closest symbols considered ($j_{max},j_{max}\leq{N}_{C}$)  is determined based on the position of $\hat{y}_{K}$ relative to $\hat{s}^{(1)}$ as indicated in Fig. \ref{f:enum}. Here ($j_{max}$) also sets the maximal child number for a particular node. For an example, based on the magnitude of $\Re|\hat{y}_{K}-\hat{s}^{(1)}|$ and $\Im|\hat{y}_{K}-\hat{s}^{(1)}|$ the square containing $\hat{y}_{K}$ relative to $\hat{s}^{(1)}$ can be identified, which sets $j_{max}$. 
 Using this, the candidate symbols corresponding to unreliable bits can be selected.  In the example of Fig. \ref{f:enum}, due to the position of $\hat{y}_{K}$ in the square highlighted in brown in between $\hat s^{(1)}$ and  $\hat s^{(2)}$ the $2^{nd}$ bit cannot be determined with high reliability and both symbols $\hat s^{(1)}$ with bit mapping $10$ and $\hat s^{(2)}$ with bit mapping $11$ need to be considered. 
 Here, $\hat s^{(3)}$ with bit mapping $00$ does not need to be considered since the $1^{st}$ bit is assumed to be determined with high reliability as $1$ if $\hat{y}_{K}$ is observed inside the region of the square highlighted in brown as in Fig. \ref{f:enum}.  Then $S_{K,1}=\hat s^{(1)}$ and $S_{K,2}=\hat s^{(2)}$ and the tree traversal proceeds to the next layer $K-1$ with these two selected candidates. The maximum number of selected candidates is $N_C$, which determines the nodes of each layer as depicted in Fig. \ref{f:enum}.
 \begin{figure}[!h]
	\centering
	\includegraphics[width=0.85\columnwidth]{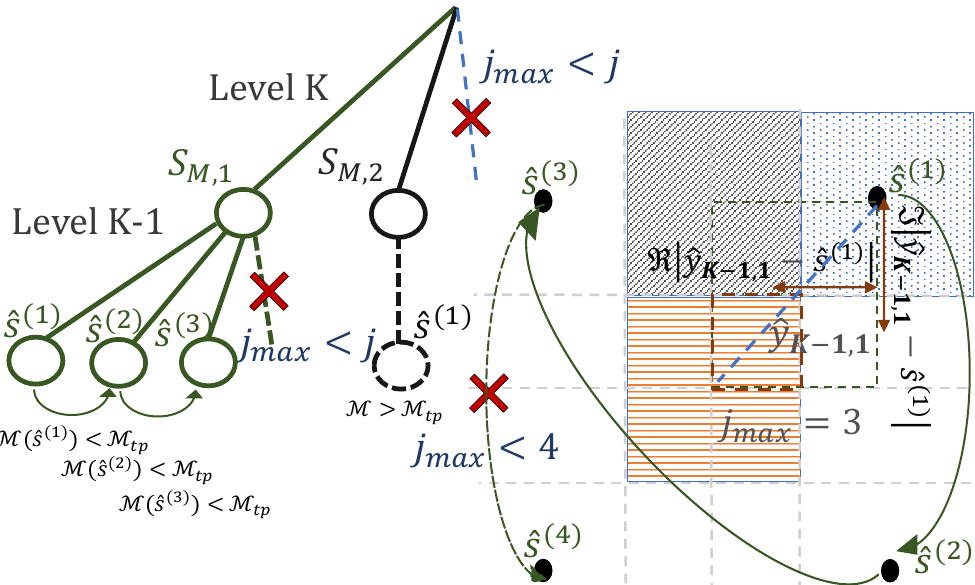}
	
	\caption{Tree  traversal and candidate update example for $l=K-1,{N}_{C}=4$.}
	\vspace{-12pt}
	\label{f:prune2}
\end{figure}

 \noindent \textit{Tree traversal and candidate update:}
As illustrated in Fig. \ref{f:prune2} in the next layer, initially, the first child node of $S_{K,1}$ and $S_{K,2}$ are expanded. This is explained in steps 12-15 in the pseudocode of algorithm \ref{pc:sortfreeARE} for a layer $l$ in general. Then, the metric of this node is compared with a threshold $\mathcal{M}_{tp}$. Similarly, the metric of the first child nodes from all the parent nodes are compared with the threshold $\mathcal{M}_{tp}$. Then, the same procedure continues with the second child node and up to $j_{max}$.
 
Fig. \ref{f:prune2} considers an example of the \textit{tree traversal and candidate update} procedure of Algorithm \ref{pc:sortfreeARE}. Fig. \ref{f:prune2} illustrates the comparison $\mathcal{M}<\mathcal{M}_{tp}$ (Step 23 of Algorithm \ref{pc:sortfreeARE}) which prunes candidate solutions with large $\mathcal{M}_n$ while avoiding any sorting operations. The candidate $\begin{bmatrix}\mathbf{\hat{S}}_{K,2} & \hat{s}^{(1)}\end{bmatrix}^{T}$ is pruned in the example illustrated in Fig. \ref{f:prune2}, since it is assumed that the corresponding $\mathcal{M}_{n}>\mathcal{M}_{tp}$. The details of the \textit{tree traversal and candidate update} procedure of Algorithm \ref{pc:sortfreeARE} are explained in steps 21-35. 
\begin{algorithm}[h] 
	\caption{ Pseudocode for The \systemname Algorithm}
	\label{pc:sortfreeARE}
	{\fontsize{8.5}{8.5}\selectfont
	\begin{algorithmic}[1]
	
	\STATE \textbf{Inputs:}$\mathbf{Q}$, $\mathbf{R}$, $\mathbf{y}$,$K$, $|\mathcal{O}|$, ${N}_{C}$, $\sigma$, $\Delta_d$, $L_T$, $\lambda$
	\STATE $\mathbf{\tilde{y}} \leftarrow \mathbf{Q}^H\mathbf{y}$
	\STATE $l \leftarrow K$ where $l$ denotes the current level
	\STATE $\mathbf{S}$ is the $K\times {N}_{C}$ list of candidate symbols
	\STATE $\mathbf{d}$ is the ${N}_{C}$ metrics of the candidate symbol vectors, initialized to zero values
	\STATE $L(b)$ is the reliability estimate of the bit $b$ 
	\STATE $\tilde{N}_{C}\leftarrow 1$ \COMMENT{$\tilde{N}_{C}$ ($\tilde{N}_{C}\leq{N}_{C})$ is the number of selected candidates}
	\STATE $\mathcal{M}$ is a buffer to contain $\tilde{N}_{C}$ metric values
	\WHILE{$l > 0$}
	\STATE{$\mathcal{M}= \emptyset$}
	  \FOR{$n\gets1,\tilde{N}_{C}$}
		 \STATE{$\hat{y}_{l,n}=\frac{\tilde{y}_{l}-\sum_{k=l+1}^{K}R_{lk}S_{kn}}{R_{l,l}}, \forall n= 1,..,\tilde{N}_C$}
		 	\STATE{Identify $1^{st}$ closest constellation symbol relative to $\hat{y}_{l,n}$ (e.g., $\hat{s}^{(1)}=\lfloor\hat{y}_{l,n}\rceil$) and initialize the symbol index $j_n$ based on relative position as $j_n=1$} 
		 		\STATE{Based on the region of $\hat{y}_{l,n}$, determine the number of closest constellation points considered ($j_{n_{max}}$,$j_{n_{max}}<N_{C}$) }\COMMENT{This can be achieved by considering  $\Re|\hat{y}_{l,n}-\hat{s}^{(1)}|$ and $\Im|\hat{y}_{l,n}-\hat{s}^{(1)}|$ as in the example of Fig. 1 }
		\STATE{$\mathcal{M}_n=\frac{|\hat{y}_{l,n}-\hat{s}^{(1)}|^2 R^2_{l,l}-\lambda^2|\hat{s}^{(1)}|^2}{\sigma^2}+d_n$} 
		\STATE{$\mathbf{\hat{S}}_{l:K,n}=\begin{bmatrix}\mathbf{S}_{l+1:K,n} & \hat{s}^{(1)}\end{bmatrix}^{T}$} \COMMENT{Update $l^{th}$ row of potential candidate $\mathbf{\hat{S}}_{n}$, where $\mathbf{S}_{l+1:K,n}$ is a partial symbol vector with elements containing symbol solutions corresponding to layers $l+1$ to $K$}
		\ENDFOR
		\STATE{$\mathcal{M}_{tp}=\min(\mathbf{d})+\frac{\Delta_d R^2_{l,l}}{\sigma^2}$} 
	\STATE{$i=0$}
	\STATE{$j_{max}=\max(j_{n_{max}})$}
	 \FOR{$j\gets1,j_{max}; i < N_C$}
	    \FOR{$n\gets1,\tilde{N}_{C}; i < N_C$}
		\IF{$ \mathcal{M}_{n}< \mathcal{M}_{tp}$}
		    \STATE{$i=i+1$}
		    \STATE{$\hat d_i=\mathcal{M}_{n}$}
		     \STATE{$\mathbf{S}_i=\mathbf{\hat{S}}_n$}
			\IF{$j_n < j_{n_{max}}$}
		    \STATE{$j_n=j_n+1$}
			\STATE{Identify the next closest (e.g.,$j^{th}_n$) constellation symbol relative to $\hat{y}_{l,n}$ of selected solution, according to predefined ordering}\COMMENT{See example in Fig. \ref{f:enum} and the Pseudocode of Algorithm \ref{pc:ordering}}
		\STATE{$\mathcal{M}_{n}=\frac{|\hat{y}_{l,n}-\hat{s}^{(j_n)}|^2R^2_{l,l}-\lambda^2|\hat{s}^{(j_n)}|^2}{\sigma^2}+d_n$} \COMMENT{Add the candidate corresponding to $j^{th}_n$ constellation symbol to $\mathcal{M}$}
		\STATE{$\mathbf{\hat{S}}_{l:K,n}=\begin{bmatrix}\mathbf{\hat{S}}_{l+1:K,n} & \hat{s}^{(j_n)}\end{bmatrix}^{T}$}
		\ENDIF
		\ENDIF
		\ENDFOR
			\ENDFOR
		\IF{$i>0$}
		\STATE $\tilde{N}_{C}\leftarrow i$
		\STATE $\mathbf{d}\leftarrow \hat{\mathbf{d}}$
		\ELSE
		 \STATE{$\mathbf{d}=\mathcal{M}$}\COMMENT{In the case of $i=0$ (No selected candidates in this layer), tree traversal proceeds to the next layer with candidates selected from previous layer}
		 \STATE{$\mathbf{S}=\mathbf{\hat{S}}$}
		\ENDIF
		
	
	 \STATE{ $l \leftarrow l-1$ }
    \ENDWHILE
     \STATE{Obtain the $M\log(|\mathcal{O}|)\times {N}_{C}$ list of bit mappings ($\mathbf{X}$) corresponding to candidate symbols ($\mathbf{S}$) }
      \STATE{$d_{1}=\min(\mathbf{d})$}\COMMENT{$\mathbf{X}_1$ is the bit mapping corresponding to $d_1$ ($\mathbf{X}_1=\text{arg}\min(\mathbf{d})$)}
     \FOR{$b\gets1,K\log(|\mathcal{O}|)$}
        \STATE{$d_{min}=\min_{\mathbf{d}\in D_{b}^{\bar{x_b}}}(\mathbf{d})$}\COMMENT{Here $ D_{b}^{\bar{x_b}}$ is a subset of $\mathbf{d}$ with the corresponding bit mapping $X_{b,j} \neq X_{b,1} $}
		\STATE{$L(b)=\text{sign}(X_{b,1})\min((d_{min}-d_1),L_T)$}\COMMENT{We take the minimum of $L_T$ and $(d_{min}-d_1)$ as magnitude of LLR.}
     \ENDFOR
	\STATE{\textbf{Output:}$L(b)$}
	
	\end{algorithmic}}
\end{algorithm}
 These comparisons, which can be performed in parallel, prune potential candidate solutions with large distance metrics at an early stage, resulting in efficient tree traversal. The pruning metric $\mathcal{M}_{tp}$ is determined as $\mathcal{M}_{tp}=\min(\mathbf{d})+\frac{\Delta_d R^2_{l,l}}{\sigma^2}$, where $\Delta_d$ depends on $d_{QAM}$ of the constellation (See Fig. \ref{f:enum}). In the evaluations, $\Delta_d=\frac{N_C+1}{8}d^{2}_{QAM}$, which seemed to provide a good performance/complexity tradeoff and was chosen to be slightly larger than the radius of the circle  ($\frac{1}{2}d^{2}_{QAM}$) which includes the $N_C=4$ constellation points closest to the origin. Here  $\mathcal{M}_{tp}$ closely follows the $N_{C}^{th}$ minimum distance. Therefore, the probability of excluding a promising candidate with a minimum distance metric is low. \textcolor{black}{Further theoretical analysis can be performed  to link $\Delta_d$ to the probabilities of detection. In particular}
 \begin{align}
 P[\hat{\mathbf{s}}\neq\mathbf{s}^t]&\leq  P[\hat{\mathbf{s}}_{ML}\neq\mathbf{s}^t]+P[\mathbf{s}^t\notin\mathcal{S}]\\
 P[\mathbf{s}^t\notin\mathcal{S}]&=1-\prod_{l=1}^{K}P[|w_l|\leq\sqrt{\Delta_d}]\approx1-\prod_{l=1}^{K}(1-e^{-\frac{\Delta_d|R_{l,l}|^2}{\sigma^2}})
\end{align}
\textcolor{black}{where $P[\mathbf{s}^t\notin\mathcal{S}]$ is the probability of the transmitted symbol vector not being included in the subset $\mathcal{S}$ searched by the \systemname, $P[\hat{\mathbf{s}}_{ML}\neq\mathbf{s}^t]$ is the maximum-likelihood error and $w_l=n_l/R_{l,l}$ is assumed to be Gaussian distributed with zero mean and variance $\sigma^2/|R_{l,l}|^2$. For the considered $\Delta_d$ value, it can be seen that $P[\mathbf{s}^t\notin\mathcal{S}]$ is less than $10\%$ for all the SNRs considered in Section \ref{s:Eval} and approaches zero with increasing SNR.}

 
\noindent \textit{Reliability estimate computation:} After following the above steps for the $M$ layers, the reliability estimate $L(b)$ of bit $b$ is obtained in the steps 45-50. In particular, first the minimum $\mathbf{d}$ value $d_1$ and its corresponding bit mapping $\mathbf{X}_1$ is identified. Then, bits with low reliability and the corresponding distance metrics ($d_{min}$) are identified in Step 48. Finally, the LLR estimate $L(b)$ is computed based on the magnitude of $d_{min}-d_1$, well approximating the LLR calculation in Eq. \ref{eq:LLR}.

\noindent \textbf{Complexity Analysis:} Step 2 of the Pseudocode in algorithm 2 requires $4MK$ real multiplications. Then, step 12 of the Pseudocode requires a maximum of $2K(K+2)2N_C$ multiplications for the $K$ layers. Steps 15 and 24 require a maximum of $12KN_C$ multiplications. Therefore, the total maximum complexity per received vector (e.g., when the maximum of $N_C$ candidate symbols are selected in all layers.) in real multiplications  is
\begin{equation}
4MK+2{K(K+2)}N_C+12KN_C.
\label{eq:comp}
 \end{equation} 
 Since $N_C \ll K$, \systemname has a time complexity of $O(MK)$ and a space complexity of $O(K^2)$.
\section{Simulation Evaluations}
\label{s:Eval}
In this Section, the performance/complexity tradeoff of \systemname is compared with the MMSE detector and related non-linear detectors by link-level simulations that employ  Rayleigh fading and 3GPP-specified channel models. For the 3GPP CDL-B  channel model \cite{ts38.901}, the UEs are assumed to be distributed randomly within a 60-degree angle from the base station. The carrier frequency was 3.5 GHz, the RMS of the channel delay spread was 300 nS, and the subcarrier spacing was set to 15kHz. LDPC coding is employed as in 3GPP standards.

\begin{figure}[!ht]	
	\centering
	\includegraphics[width=\columnwidth]{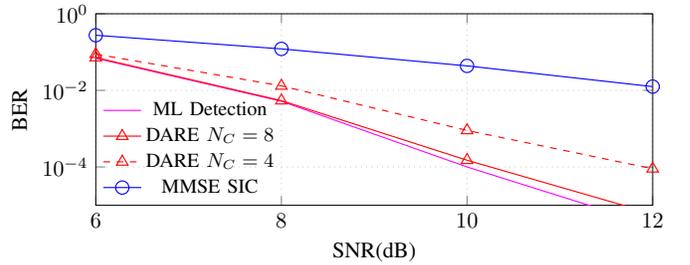}
	\caption{12-antenna base station supporting 12 users in a Rayleigh fading channel. 16 QAM, uncoded transmissions are employed.}
	\label{f:BER}
\end{figure}

\begin{figure}[!ht]	
	\centering
	\includegraphics[width=\columnwidth]{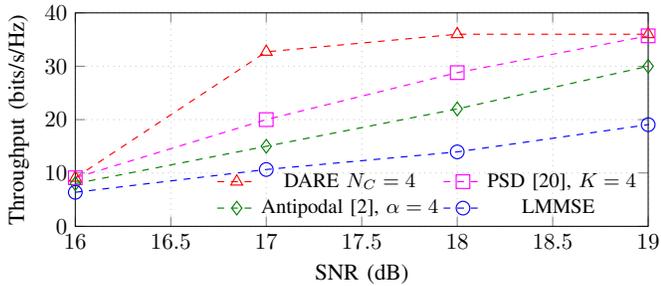}
	\caption{12-antenna base station supporting 12 users in a Rayleigh fading channel. 16 QAM, 0.75 rate is assumed. }
	\label{f:MIMO12x12}
 \vspace{-12pt}
\end{figure}
In Fig. \ref{f:BER}, we compare the Bit-Error-Rate (BER) of \systemname, ML detection, and MMSE with successive interference cancellation (SIC). \systemname approaches ML detection performance when $N_C=8$, significantly outperforming MMSE SIC with comparable complexity requirements. 
To obtain insight into system performance, in Fig. \ref{f:MIMO12x12}, we compare the achievable throughput (\cite{STS,Antipodal}) of \systemname, Probabilistic Searching Decoding (PSD), \cite{PSD}, Antipodal \cite{Antipodal} and linear MMSE (LMMSE) detection for a $12\times12$ MIMO-OFDM system where each subchannel between transmit and receiver antenna is modeled by a four tap i.i.d Rayleigh channel. As shown, \systemname can provide significant gains (e.g., $>$ 3dB) when $N=M$, while the maximum complexity is  $2\times$ in comparison to LMMSE. Obtaining soft information from MMSE SIC is computationally expensive than \systemname while the performance is poor.

\begin{figure}[!ht]
\begin{subfigure}{.95\columnwidth}
\centering
\includegraphics[width=\columnwidth]{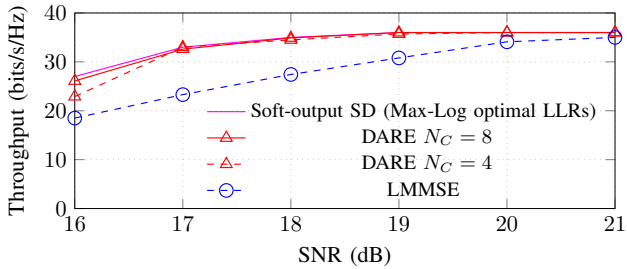}
\caption{Throughput comparison}
\label{f:MIMO64x12a}
\end{subfigure}
\begin{subfigure}{.95\columnwidth}
\centering
\includegraphics[width=\columnwidth]{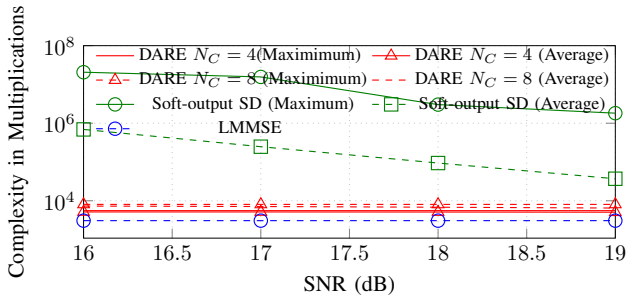}
\caption{Per vector complexity}
\label{f:MIMO64x12b}
\end{subfigure}
\caption{64-antenna base station supporting 12 users in the CDL-B channel. 16 QAM, 0.75 rate is assumed.
}
\label{f:MIMO64x12}
\vspace{-12pt}
\end{figure}
In Fig. \ref{f:MIMO64x12a}, we compare the achievable throughput of \systemname with $N_C=4$ and $8$, the soft-output SD \cite{STS} which can provide optimal soft-output detection performance and LMMSE, for a $64\times12$ MIMO-OFDM systems modeled by a 3GPP-CDL-B channel.  As shown, \systemname can achieve the performance of optimal soft-output detection, achieving throughput gains of $40\%$ in comparison to LMMSE. As shown in Fig. \ref{f:MIMO64x12b}, the complexity of \systemname is at least two orders of magnitude lower than the soft-output SD and less than $2\times$ that of LMMSE.
\begin{figure}[!ht]	
	\centering
	\includegraphics[width=\columnwidth]{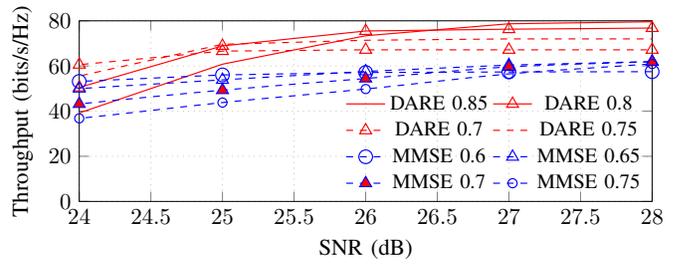}
	\caption{64-antenna base station supporting 16 users in the CDL-B channel. 64 QAM with multiple code rates is assumed.}
	\label{f:MIMO64x16a}
\end{figure}

In Fig. \ref{f:MIMO64x16a}, we compare the achievable throughput of \systemname and MMSE for a $64\times16$ MIMO-OFDM system with multiple code rates. Even when the code rate that maximizes the throughput is selected at each SNR, to model perfect adaptive modulation and coding, \systemname can provide throughput gains of $38\%$. 
\systemname can still provide throughput gains in massive MIMO scenarios due to the spatial correlation of receiver antennas, which is taken into account by the CDL-B channel. 



\begin{figure}[!ht]
\begin{subfigure}{.95\columnwidth}
\centering
\includegraphics[width=\columnwidth]{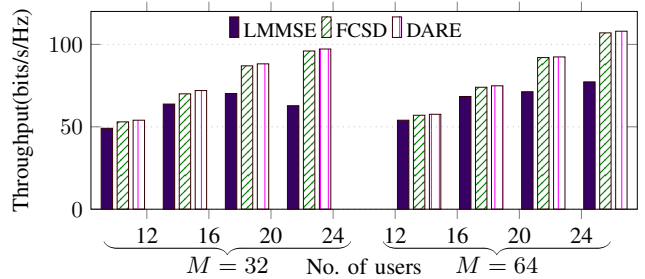}
\caption{Throughput comparison}
\label{f:MIMOCDL}
\end{subfigure}
\begin{subfigure}{.95\columnwidth}
\centering
\includegraphics[width=\columnwidth]{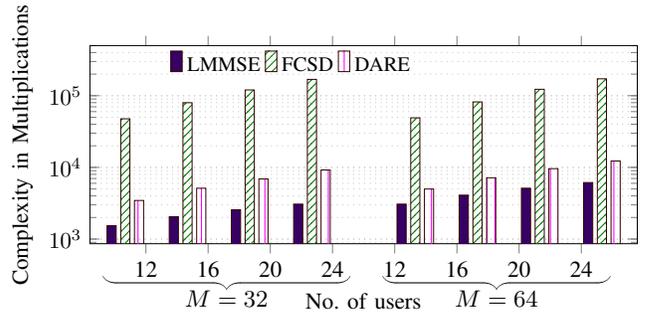}
\caption{Per vector complexity}
\label{f:compCDL}
\end{subfigure}
\caption{MU MIMO-OFDM with $M=32$ and $M=64$, with varying $K$(i.e., $K=12,16,20,24$) modeled by a 3GPP CDL-B channel at an SNR of 24 dB. The employed modulation order is chosen from 4, 16, 64, and the code rate from 1/2, 2/3, 3/4, and 5/6 to maximize throughput.}
\vspace{-12pt}
\end{figure}

In Fig. \ref{f:MIMOCDL}, we compare the achievable throughput of \systemname, Fixed Complexity Sphere Decoder (FCSD) \cite{newsoftfsd,SFSD} and MMSE detector for a MU MIMO-OFDM with $M=32$ and $M=64$, with varying $K$(i.e., $K=12,16,20,24$) modeled by a 3GPP CDL-B channel at an SNR of 24 dB. 
As shown in Fig. \ref{f:MIMOCDL}, \systemname can provide up to $40\%$ gain in throughput in comparison to MMSE and achieve better throughput with half the number of base station antennas. As shown in Fig. \ref{f:compCDL}, the complexity of \systemname is two orders of magnitude lower than FCSD while achieving similar throughput.

\section{Conclusions}

This work introduced  (\systemname): a highly efficient ultra-low-complexity non-linear detection scheme that can significantly outperform linear detection, providing substantial throughput gains (e.g., $40\%$) with a complexity that is only 1.5x. Furthermore, \systemname can achieve better throughput than linear detection using half the base station antennas, significantly reducing the base station power consumption. Due to these gains, \systemname becomes a strong candidate for future power-efficient MU-MIMO developments.

\bibliographystyle{IEEEtran}
\bibliography{reference}

\end{document}